\documentclass[10pt,conference]{IEEEtran}
\IEEEoverridecommandlockouts
\usepackage{graphicx}
\usepackage{lscape}
\usepackage{hanging}
\usepackage{amsfonts}
\usepackage{amssymb}
\usepackage{amsmath}
\usepackage{subfigure}
\usepackage{rotating}
\usepackage{latexsym}
\usepackage{epsf}
\usepackage{epsfig} 
\usepackage{multirow}

\begin{document}

\title{\textit{PROTECT}: \textit{Pro}ximity-based \textit{T}rust-advisor using \textit{E}n\textit{c}oun\textit{t}ers for Mobile Societies}
\author{\IEEEauthorblockN{Udayan Kumar, Gautam Thakur, Ahmed Helmy}
\IEEEauthorblockA{\\Department of Computer and Information Science and Engineering, University of Florida
\\Email: \{ukumar, gsthakur, helmy\}@cise.ufl.edu} }

\maketitle
\begin{abstract}
 Many interactions between network users rely on trust, which is becoming particularly important given the security breaches in the Internet today. These problems are further exacerbated by the dynamics in wireless mobile networks. In this paper we address the issue of trust advisory and establishment in mobile networks, with application to ad hoc networks, including DTNs. We utilize encounters in mobile societies in novel ways, noticing that mobility provides opportunities to build proximity, location and similarity based trust. Four new trust advisor filters are introduced - including encounter frequency, duration, behavior vectors and behavior matrices - and evaluated over an extensive set of real-world traces collected from a major university. Two sets of statistical analyses are performed; the first examines the underlying encounter relationships in mobile societies, and the second evaluates DTN routing in mobile peer-to-peer networks using trust and selfishness models. We find that for the analyzed trace,  trust filters are stable in terms of growth with time (3 filters have close to  90\% overlap of users over a period of 9 weeks) and the results produced by different filters are noticeably different. In our analysis for trust and selfishness model, our trust filters largely undo the effect of selfishness on the unreachability in a network. Thus improving the connectivity in a network with selfish nodes. 
 
We hope that our initial promising results open the door for further research on proximity-based trust.

%The selfishness increases the unreachability in the network by $33\%$ and the presence of trust as generated by trust filters reduced it to $11\%$ of the network with no selfish nodes. 

\end{abstract}
\section{Introduction}
Most interactions between people take into consideration, and in many cases rely on, prior established trust. In the context of interactions over computer networks trust establishment is particularly important and challenging, given the identity theft, fraud and security breaches common in the Internet today. These problems are further exacerbated by the uncertainty and dynamics in wireless mobile networks. Furthermore, in infrastructure-less peer-to-peer mobile networks, such as ad hoc, delay tolerant (DTNs) or sensor networks, cooperation, and subsequently trust is imperative to the construction and operation of the network. Without cooperation and trust these networks would fail. With trust, several potential applications can be enabled including mobile social networking, building groups and communities of interest, localized alert and emergency notification, context-aware and similarity-based networking~\cite{profilecast}, to name a few.

This study addresses the issue of trust advisory and establishment in mobile networks, with application to infrastructure-less wireless ad hoc networks, including DTNs. We utilize encounters in mobile societies in novel ways, noticing that mobility provides opportunities to build proximity, location and similarity based trust. Each mobile device running our trust application shall keep simple history of other devices encountered (e.g., using existing neighbor discovery) and (if applicable) the location of the encounters (e.g., using GPS or access point association history). Four new so-called trust advisory filters are introduced to analyze the encounter and location history, including: 1. encounter frequency ($FE$) based filter, 2. encounter duration ($DE$) based filter, 3. encounter location based behavior-vector ($BV$)  filter  based on - Count and  Duration, and 4. location preference based behavior-matrix ($BM$) filter. Each trust advisory filter provides a score reflecting the level of trust identified by our system, to aid the users in identifying potential trustworthy nodes in the network. The filters calculations are done in a fully distributed fashion that obviates the need for any server or trusted third party. Ultimately, an application with these filters shall provide the user with an option to choose trustworthy nodes in coordination with personal preferences, location priorities, contextual information, or encounter-based keys, as explained in Section~\ref{trust_filter}.

The trust advisory filters are analyzed and evaluated over an extensive set of real-world traces collected from a major university, including over 35,000 users and spanning 12 weeks. Two sets of statistical analyses are performed. The first examines the underlying encounter relationships in mobile societies, including distribution of encounter frequency, duration, similarity scores based on location vectors and matrices, and the differences (and similarities) between the various proposed filters. The second evaluates DTN routing in mobile peer-to-peer networks using trust and selfishness models, where epidemic routing is used to analyze the characteristics of the dynamic encounter graph.

We find that for the analyzed trace, three filters ($DE$, $FE$, and $BV$-Count) create trust lists that remain consistent (with 90\% overlap) for more than 9 weeks. Thus the results from these filters are stable.  Initially, all the filters were expected to produce similar results in terms of recommended trusted nodes, but the results showed noticeable differences in the selection. Surprisingly, despite these differences the overall performance of epidemic routing based on the different filters (or even random selection) were found to be similar. This suggests that our methods for trust advisory can indeed provide the users with trust options without the sacrifice of performance in the context of DTNs. We note from the trust and selfishness modeling that having trusted nodes as recommended by the trust filters largely eliminates the effect of selfishness in the network. The use of trust filters decreases the unreachability in the selfish network (from 0.146 at selfishness probability ($S$)=0.8 and trust ($T$)=40\% to 0.122 using FE filter when 0.109 is the unreachability in network with no selfishness).

To sum, the main contributions of this paper include: \linebreak 1- The proposal of a trust advisory framework that utilizes mobile encounter characteristics, evolution and locations, \linebreak 2- The introduction of four specific trust advisory filters based on encounter frequency, spatio-temporal statistics and similarity scores. The filters were evaluated using extensive mobile network traces from a university campus, compared and contrasted. 3- Performance of epidemic routing, in the context of DTN, was analyzed for the various trust advisory filters in conjunction with selfishness models. It was found that our trust advisors can indeed be utilized in DTN without adversely affecting the network performance.

We note that our framework of establishing trust incrementally over a time window can be extended to include encounter-based keys~\cite{gangs,smile},  using out of band exchange and to scenarios where trust evolves - strengthens or dissolves - over time. The studies conducted here focus on trust establishment to aid network connectivity and interaction, after which incentive~\cite{sprite,nuglets} and reputation~\cite{reputation} systems may be used to police, monitor and reward transfers. Our work is thus complimentary to, and can be integrated with, related work in these areas. Our future work shall investigate this direction further, in addition to analyzing other real-world traces from other contexts, and implementing, deploying and monitoring the trust advisory application in a test-bed.

Following is an outline for the rest of this document, Section~\ref{rel_work} discusses related work. Section~\ref{enct_trust} provides overall motivation for using encounters. Section~\ref{trust_filter} introduces the trust advisory filters. Section~\ref{stat_ana} provides the analysis of the filters using mobile network traces. Section~\ref{epidemic} analyzes performance of epidemic routing using trust and selfishness. Section~\ref{future} provides future work, applications and conclusion.

\section{Related Work} \label{rel_work}

In this section, we discuss some of the recent work being done to establish trust and cooperation in the delay tolerant networks, which is a critical aspect in networks to communicate and deliver information. Several researchers have proposed novel approaches to establish trust and cooperation in mobile ad hoc and delay tolerant networks using credit and reputation based schemes, incentive and trust based schemes, game theory and efficient routing based algorithms. 

The reputation based scheme targets better peer selection based on previous interactions records and transfer, thus providing trust ratings and cooperation to nodes in a mobile ad-hoc network. In \cite{reputation} a node detects misbehavior locally by observation and use of second-hand information, in \cite{robustreputation} propose a fully distributed reputation system that can cope with false disseminated information where each node maintains a reputation rating and a trust rating about other nodes. In \cite{Pavan03cooperationin,Altman04non-cooperativeforwarding,1032153} analysis of rewards provisions and punishment is conducted based game theoretic approaches to provide incentives for message delivery. In \cite{conf/netcoop/Altman09}, authors derived performance and optimization statistics to measures the success in delivery probability for a message covering both cooperative and non cooperative scenarios. The study in  \cite{coopdtn} analyzed the effect of cooperation on three different routing algorithms. The authors investigated the performance of epidemic, two-hop relaying and binary spray and wait routing to model a node's cooperation probability to either drop a message or to forward the message to next hop level. However, these works focus mainly on limited size topology implementation, for only in particular scenarios. The routing and message delivery work do not address social trust development and cooperation in DTNs, which we focus on in this study based on historical events and incremental build-up of social parameters based on node encounters.

The incentive based credit schemes rank the trust on neighboring nodes, in \cite{4351689}, authors proposed a game-theoretic model to discourage selfish behavior and stimulates the cooperation by leveraging Nash equilibria with socially optimal behavior, while in \cite{sprite}, authors proposed a pricing mechanism to give credits to the nodes who participate in message forwarding mechanism. However, in these, the cooperation is developed based on the number of messages transfered rather than socially acceptable criteria of encounter analysis. 

Message delivery mechanisms in DTN require node collaboration. However, in real deployment some nodes may refuse to cooperate. Such selfish nodes (or free riders)~\cite{And02simulation-basedanalysis} could exploit the services by receiving the messages from other nodes but refusing to deliver it to next hop level. An analytical model that builds the concept of trust is discussed in  \cite{Marti00mitigatingrouting,proptrust}. The authors show trust supports cooperation and is heavily based on the interactions and bonds that govern behavior in ad hoc and opportunistic scenarios.

Other approaches discussed in \cite{spate,gangs,seeing} propose explicit authentication mechanism to generate trust and cooperation in network. These approaches are better modeled for small group deployments~\cite{spate} and requires to exchange public keys and the installation of the private key on the users device~\cite{gangs}. They do not address some of issues of selfish nodes, scalability and overhead incurred due to the key exchange and global optimization.

%
%All these works provide an important analysis that trust and cooperation is one of cornerstones of message delivery in delay tolerant networks. In this paper, we investigate frequency and duration of node's encounter in building trust and cooperation in the network and provide analysis for routing message based on these metrics. The proposed work can be used as an advisory to the aforementioned related works. 

%\subsection{Definitions}
%Below are the definitions of some of the terms used throughout the paper.
%
%\textbf{Proximity:}  When a device being in the range of atleast one wireless communicating device, such that information can be exchanged between them, we can say that these devices are in proximity of each other. Many of the wireless technologies like WiFI (802.11) and Bluetooth allow for device discovery. The device discovery shows the devices whose signal the device can sense i.e. the devices in proximity.  
%
%\textbf{Encounter:} The event when two devices are in proximity is called Encounter.

\section{Traces Used}
For the purpose of this study we have used 3 month long (Sep to Nov 2007) Wireless LAN (WLAN) traces coming from University of Florida, Gainesville. More than 35,000 users who accessed the WLAN during this period have been considered in the study. Total number of Access Points is over 730. The information provided in the traces is anonymized. Tab.~\ref{real_trace} shows a sample of the traces used in this work.

\begin{table}[htp]
\begin{center}
\tiny
\begin{tabular}{c c c c}
\hline MAC & Start Time & Duration(sec) & AP/Location \\
\hline 00:11:22:33:44:55 &   01 Jun 2008 21:00:51 GMT &     3000secs &   CS\_buildingAP1\\
11:22:33:44:55:66 &   01 Jun 2008 21:01:30 GMT &       10secs &   ECE\_buildingAP2 \\
01:02:03:04:05:06 &   01 Jun 2008 22:11:00 GMT &       200secs &   MSL\_buildingAP1 \\
10:20:30:40:50:60 &   01 Jun 2008 22:15:30 GMT &       600secs &   MACA\_buildingAP1 \\
11:22:33:44:55:66 &   01 Jun 2008 22:23:10 GMT &       180secs &   CS\_buildingAP3 \\
\hline \\
\end{tabular}\end{center} 
\caption{Sample WLAN trace}
\label{real_trace}
\end{table}

\section{Encounter/Proximity based Trust} \label {enct_trust}

The foremost purpose of this work is to encourage interactions in the mobile societies and establish network connectivity in the context of ad-hoc and DTN networks. Trust can be the foundation of cooperation in networks in general and particularly in the infrastructure-less networks. In our context trust means that a user is: 1. Willing to interact through the network with the trusted node, and 2. In the context of DTN, ready to accept a message for the trusted user and genuinely attempt to route it.  To develop trust between a pair of users, we  leverage interactions between users based on proximity (when the devices come within radio range) and encounter information. Several properties of nodal encounter behavior have been investigated in~\cite{nodal}.
%As suggested by numerous social sciences studies on Homophily~\cite{homophily}, the more the encounters the higher is the probability of interactions and trust.  

The primary reasons for choosing encounters and proximity as measure to generate trust comes from the work on Homophily~\cite{homophily}. The principle of Homophily states that people with similar interest meet and interact often. Trusting  frequently encountered users would mean trusting people similar to oneself (like work colleagues or classmates). This trust can have social incentives too. The second reason for choosing proximity/encounter for trust is that because the users are within the radio range of each other (for Bluetooth it is $\sim$15m) they can exchange out-of-band information. This out-of-band exchange can enable a user to know more about the other user and can also be used to set-up cryptographically strong secrets between them~\cite{gangs}. This feature of out-of-band information exchange is not possible on wired networks (as the two terminals may be geographically far apart) but can be easily leveraged by mobile peer to peer wireless networks.

The challenge is to find methods that can successfully discover potential similarities between the users. We refer to these methods  as Trust  Advisors Filters.  In the implementation, a user would actually decide on which users to trust and the filters would serve as an advisors. Thus users would have full control over the selection of trusted users. These filters would act as the recommendation system that recommends users who are most similar to the user. In the next section we present a rich set of filters that find users most similar to the user.

\section{Trust Adviser Filters/Metrics}  \label{trust_filter}

This section presents a set of filters that capture similarity between two users. This similarity can be captured indirectly by using encounters  and measuring metrics such as frequency of encounters, duration of encounters, etc. The similarity can also be captured directly by comparing the behavior profiles/preferences with another user such as by comparing the time spend at different locations. Behavior-Matrix and Behavior-Vector based approaches discussed below belong to the class of filters that can directly measure similarity.  Following we introduce A. Frequency of Encounters, B. Duration of Encounters, C. Behavior Vector, and D. Behavior Matrix filters.

\subsection{Frequency of Encounters (FE)}
One of the basic filters to estimate similarity between a user-pair is the number of times they encounter (An encounter is defined as the event where a device is in radio range of another device such that it can discover the other device's MAC address).  The more they meet, the more similar (and are more trustworthy) they are.  On this basis, we design  the $FE$ filter that counts the frequency of the encounters of the user with all the other users. To get the trust list from $FE$ filter for a user, we sort all the encountered users by their number of encounters and select top users based on the trust ($T$) value. 

The computational complexity of $FE$ is $O(N)$ and space complexity is $O(N_u)$ per device, where N is the number of encounters and $N_u$ is the number of unique users encountered.

\subsection{Duration of Encounters (DE)}
Another fundamental filter to estimate similarity is the percentage of time spend by the user with another user. The more the time spent together by the users, the more similar they are (and are more trustworthy) likely to be. On this basis we design the $DE$ filter to keep count of the duration of time spent by a user with all the other users. From the ordered list of duration of encounters for the user, $DE$ filter selects top trusted users based on the $T$ value. 

The computational and space complexities for $DE$ are similar to those of $FE$, computation complexity of $O(N)$ and space complexity of $O(N_u)$ per device.

\subsection{Behavior Vector (BV)}
\begin{figure}
\centering
\includegraphics[scale=.5]{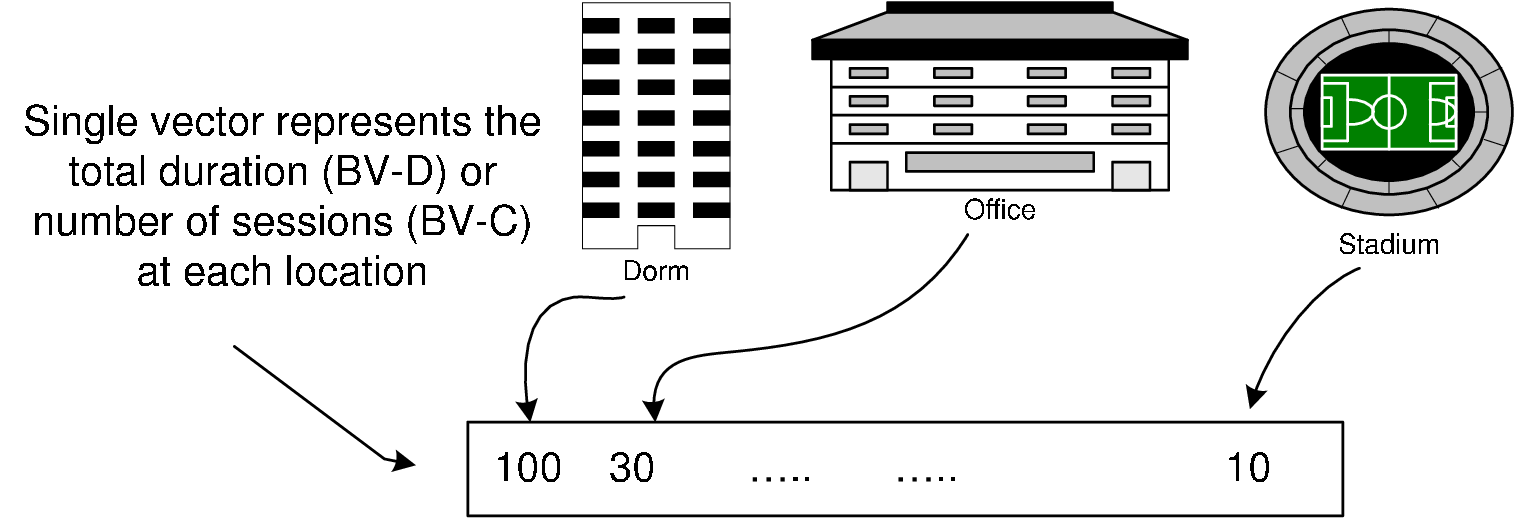}

\caption{Behavior Vector for a user}
\label{dig:PV}
\end{figure}

An important characteristic of homophily is that similar people tend to go to similar locations. To capture this characteristic, we have designed a filter that stores location preferences of a user in a single dimensional vector. Each user maintains a single vector for oneself. The columns of the vectors represent the different location visited by a user and the values stored in each cell indicate either duration ($BV-D$) or count ($BV-C$) of the sessions at that particular location, depending on the type of $BV$ used. To get similarity score between two users, they exchange the normalized BVs and compute the inner product of the two normalized BVs. The result of the inner product gives a score indicating the similarity between the two users. This score is higher if the two BV are  similar and can be zero if the users do not have single visited location in common. Fig.~\ref{dig:PV} illustrates the design of a Behavior Vector.

The space complexity of BV on each device is $O(L+N_u)$, where L is the number of unique locations visited by the user and $N_u$ is the number of unique users whose similarity scores are stored on the device. The computation complexity for computing similarity score for two BV of length L1 and L2 is $O(L1+L2)$.

\subsection{Behavioral Matrix (BM)}

\begin{figure}
\centering
\includegraphics[scale=.3]{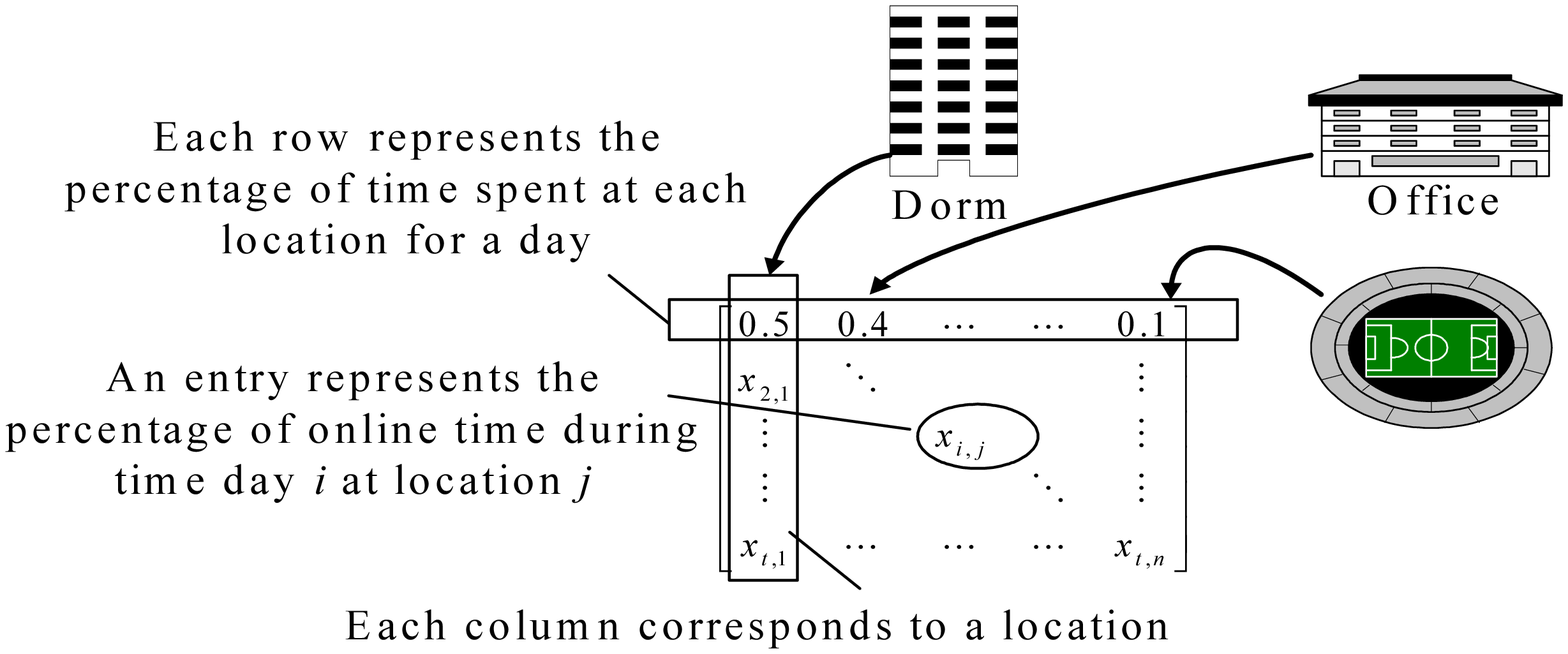}

\caption{Behavior Matrix  for a  user (from \cite{profilecast})}
\label{dig:svd}
\end{figure}

We notice that for $BV$ we maintain two vectors for duration and frequency to capture user behavior. A modification of vector into a matrix allows us to maintain a single entity to capture a spatio-temporal representation of user behavior. In this behavior matrix each column denotes a location and each row represents a single day (here the frequency is counted in terms of days). The value stored at each cell is the fraction of the on-line time spent by the user at a particular location on a particular day as shown in Fig.~\ref{dig:svd}. To get the similarity score between two users, we compare the two matrices. This similarity has been exploited by the set of Profile-Cast protocols~\cite{profilecast} and Behavior mining techniques~\cite{mineweijen}. 

To make the behavior similarity check  efficient (in terms of space and computation complexity), we use the eigen values of the behavior matrix for exchange between the two users. The eigenvalues are generated using SVD (Singular Value Decomposition). SVD is applied to a behavior matrix $M$, such that:

\begin{equation} \label{svd-eq}
M = U \cdot \Sigma \cdot V^T,
\end{equation}
where a set of {\it eigen-behavior} vectors,  $v_1, v_2, ..., v_{rank(M)}$ that summarize the important trends in the original matrix $M$ can be obtained from matrix $V$, with their corresponding weights, $w_{v_1}, w_{v_2}, ..., w_{v_{rank(V)}}$ calculated from the eigen-values in the matrix $\Sigma$.  This set of vectors is referred to as the {\it behavioral profile} of the particular user, denoted as $BP(M)$, as they summarize the important trends in user $M$'s behavioral pattern. The {\it behavioral similarity} metric between two users' association matrices $A$ and $B$ is defined based on their {\it behavioral profiles}, vectors $a_i$'s and $b_j$'s and the corresponding weights, as follows: 
\begin{equation} \label{sim-index}
Sim(BP(A),BP(B)) = \sum^{rank(A)}_{i=1} \sum^{rank(B)}_{j=1}
w_{a_i}w_{b_j}\vert a_i \cdot b_j \vert
\end{equation}
which is essentially the weighted cosine inner product between the two sets of {\it eigen-behavior} vectors.

The space complexity of $BM$ filter is $O(DL)$ per device, where L is the number of locations visited and D is the number of days included in the matrix.  Computation complexity of SVD is $O(D^2L+DL^2)$.

If we compare $FE$ and $DE$ filters with $BV$ and $BM$, we find that $BV$ and $BM$ require information exchange between two users to calculate similarity. Since a user has to depend on information provided by another user, issues of data authenticity and privacy become important along with the additional costs incurred in communication for the transfer of vectors and matrix. To maintain authenticity of the data, several techniques have been proposed that use  Trusted Platform Module (TPM)~\cite{trustedcomputing,towardstrusted}, this module maintains the integrity of the sensed data. To lower the communication costs in $BM$, one can only send top 5 to 10 eigen values as it has been shown these can capture 90\% of the user behavior~\cite{profilecast}.

%The communications cost for both BM and BV communication costs can be removed by a small modification in the algorithm.  If a user maintains a behavior matrix or vector for all the users encountered and updates the matrix or vector of a particular user whenever this user is encountered, there is no need to exchange BVs and BMs. This will increase the space complexity and may not be able to capture profiles as well as provided by the users themselves, but this variation would not require any data transfer (thus no issue of data authenticity, privacy or communication cost).

\subsection{Random Trust (RT)}
This filter as the name suggests selects $T$ percent of encountered users randomly and adds them to the trust list, we use this filter for comparison purposes. Ideally, we would never want to refer to the advise by this filter, but we have kept it here to later study the impact of each filter with respect to random selection.

We note that for each filter  the user may also choose to have a lower threshold for selecting users before they are considered by the filters. For example, in $FE$ filter the user may not want to consider users who have less than 2 encounters. However, due to lack of space we are not presenting the analysis using thresholds.

\section{Statistical Analysis of Trust Filters} \label{stat_ana}
Trust adviser filters need information about previous encounters (history of encounters) to  produce trust lists. In this section we conduct statistical analysis to investigate trust list's stability with time, similarity between the trust list generated by the filters, and encounter similarity.  This analysis can give us insight into the requirements and benefits of filters. For all the following analysis we use the WLAN traces.

\subsection{Encounter and Similarity Statistics}

To use a filter, it is necessary to know if the filter is going to give any results at all and would it be able to distinguish the users for trust advising.  In this section, we present evaluation of the proposed filters.

\subsubsection{FE}
\begin{figure}
\centering

\includegraphics[width=2.5in]{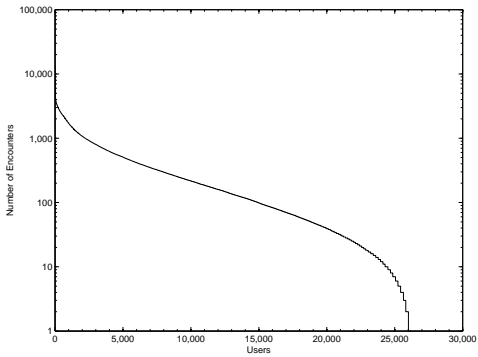} 
\caption{Encounters per user in the month of Nov 2007 (ordered in terms of Encounter Frequency)}
\label{fig:FE}
\end{figure}

Fig.~\ref{fig:FE}  illustrates encounters per user for the month of Nov 2007. The figure shows that out of the total population, 3,000 users have more than 1,000 encounters each in a month and more that 15,000 users (more than $2/3$ of the population) have over 100 encounters. This result shows that users have significant number of encounters and it justifies our use of $FE$ as a trust filter.  

\subsubsection{DE}
\begin{figure}
\centering

\includegraphics[width=2.5in]{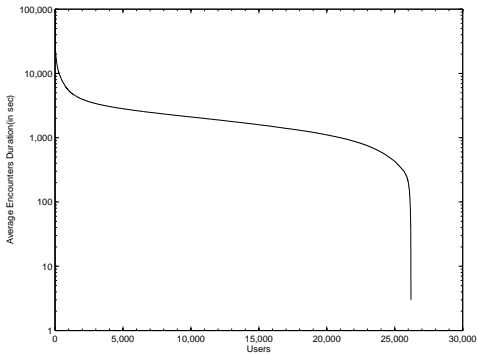} 
\caption{Average Duration of Encounter per user in the month of Nov 2007 (ordered in terms of Duration of Encounter)}
\label{fig:DE}
\end{figure}

Fig.~\ref{fig:DE} illustrates the average duration of encounters for Nov 2007. We notice that average encounter duration for more than 20,000 users is greater than 1,000 seconds. This justifies the use of DE as a trust filter.

\subsubsection{BV}
\begin{figure}
\centering
\includegraphics[width=2.5in]{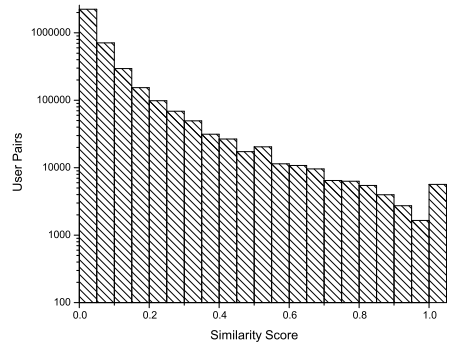} 
\caption{Similarity score using Behavior Vector-Duration($BV-D$) filter for all the encountered pairs of users in Nov 2007 }
\label{fig:BV-dura}
\end{figure}

 Fig.~\ref{fig:BV-dura} illustrates the similarity score between all the pair of users who encountered at least once in the month of Nov 2007. Large number of user pairs have low similarity (close to zero), which may mean that most of the users are not similar to each other and few user pairs have very high similarity scores. Results for $BV-C$ are similar to $BV-D$ (Fig.~\ref{fig:BV-dura}).

\subsubsection{BM}

\begin{figure}
\centering

\includegraphics[width=2.5in]{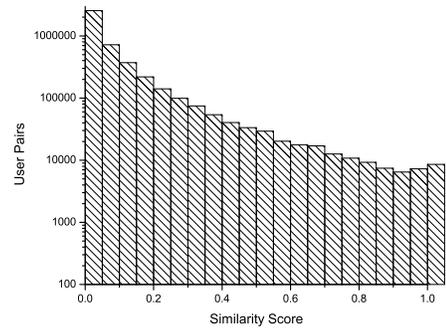} 
\caption{Similarity score using Behavior Matrix ($BM$) filter for all the encountered pairs of users in Nov 2007 }
\label{fig:BM}
\end{figure}

$BM$ similarity score like $BV$ is close to zero for most of the user pairs and is high for a very few user pairs. This is illustrated in Fig.~\ref{fig:BM}. This implies that $BM$ scores would be able to distinguish between the users. 

\subsection{Stability Analysis}

\begin{figure*}
  \begin{center}
\renewcommand{\thesubfigure}{\Alph{subfigure}.}

    \centering
    
      \subfigure[Duration of Encounter($DE$)]{\epsfig{file=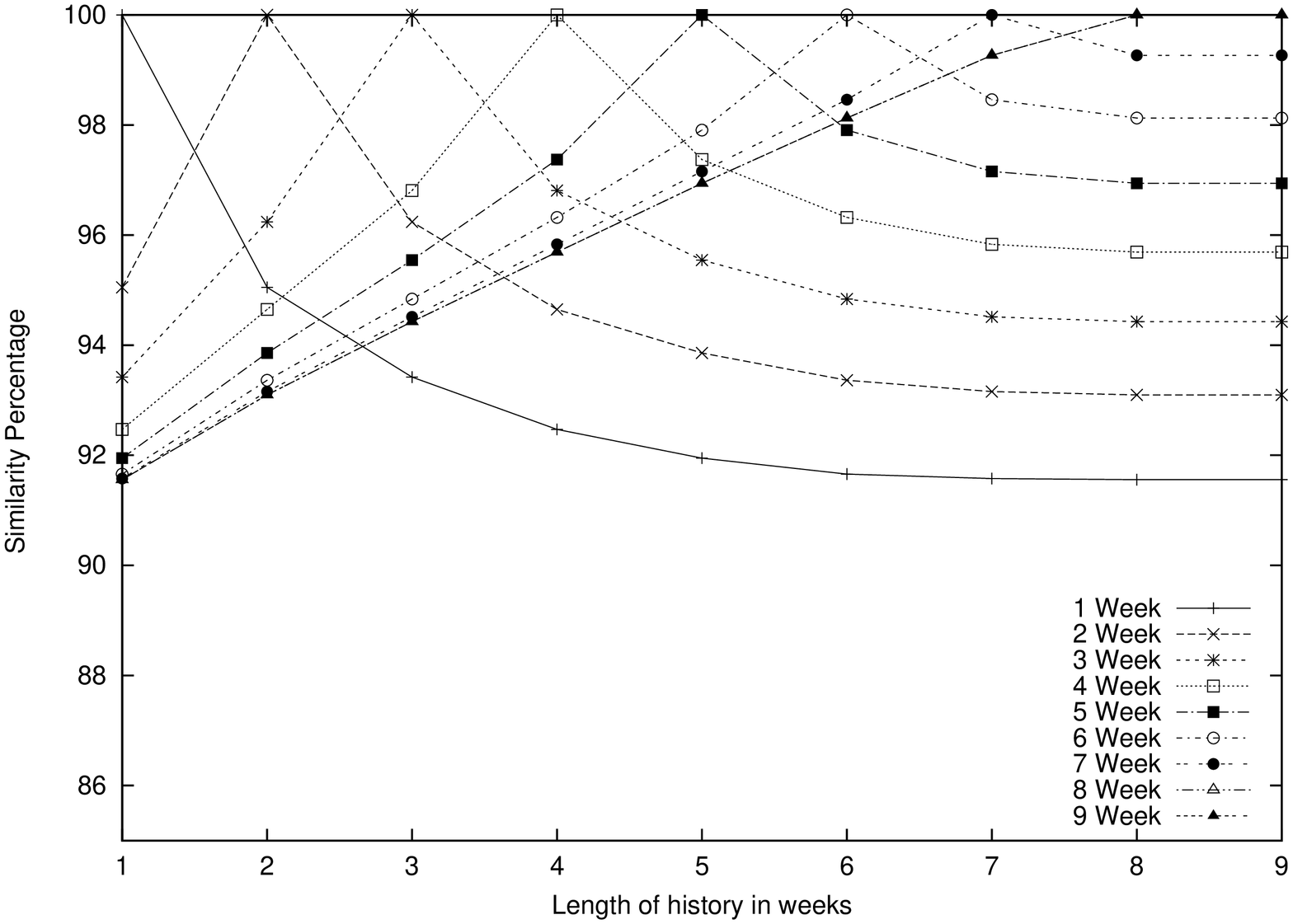, scale=0.22,angle=270}} 
      \subfigure[Frequency of Encounter ($FE$)]{\epsfig{file=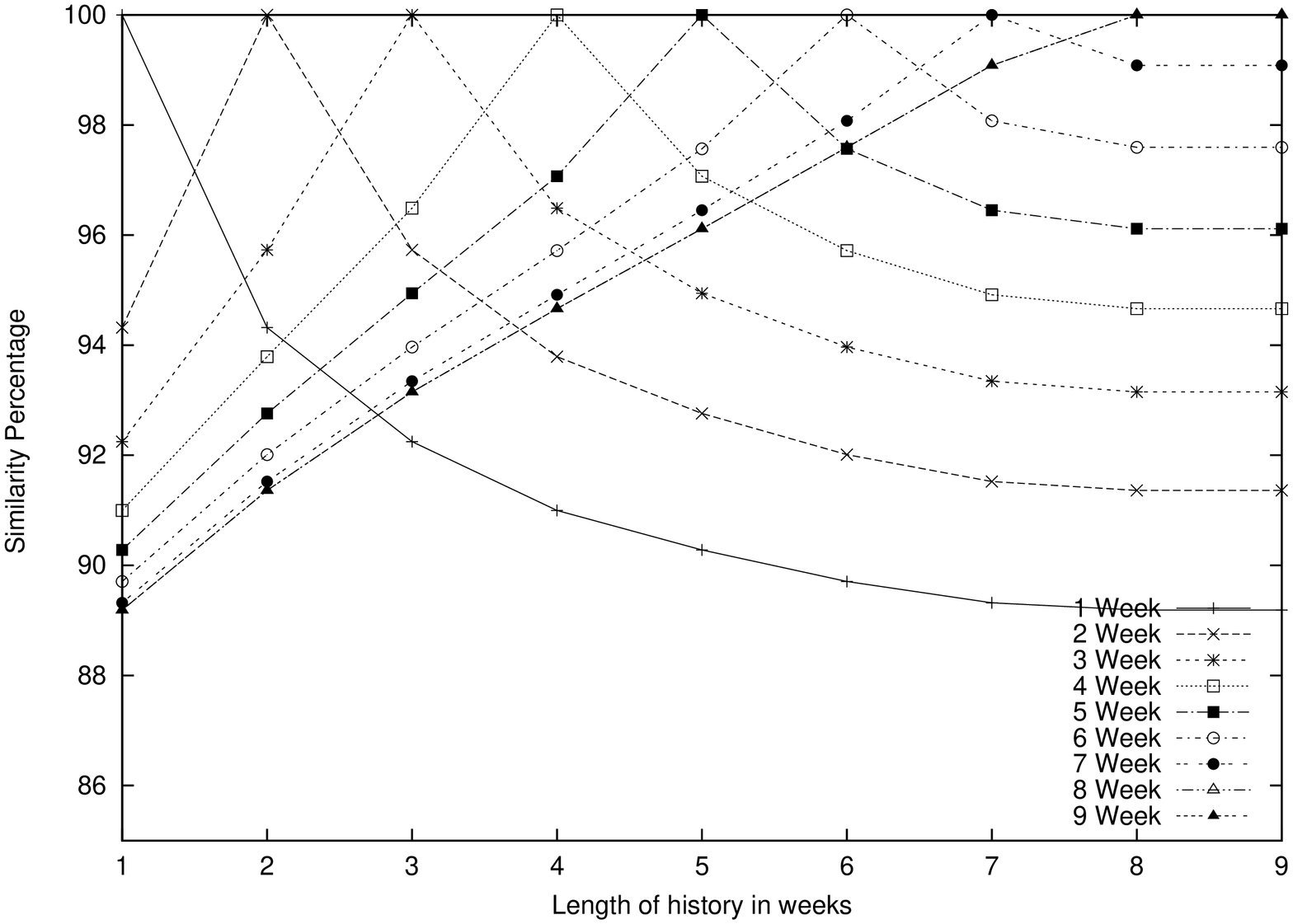, scale=0.22,angle=270}} 
      \subfigure[Behavior Vector using ($BV-C$)]{\epsfig{file=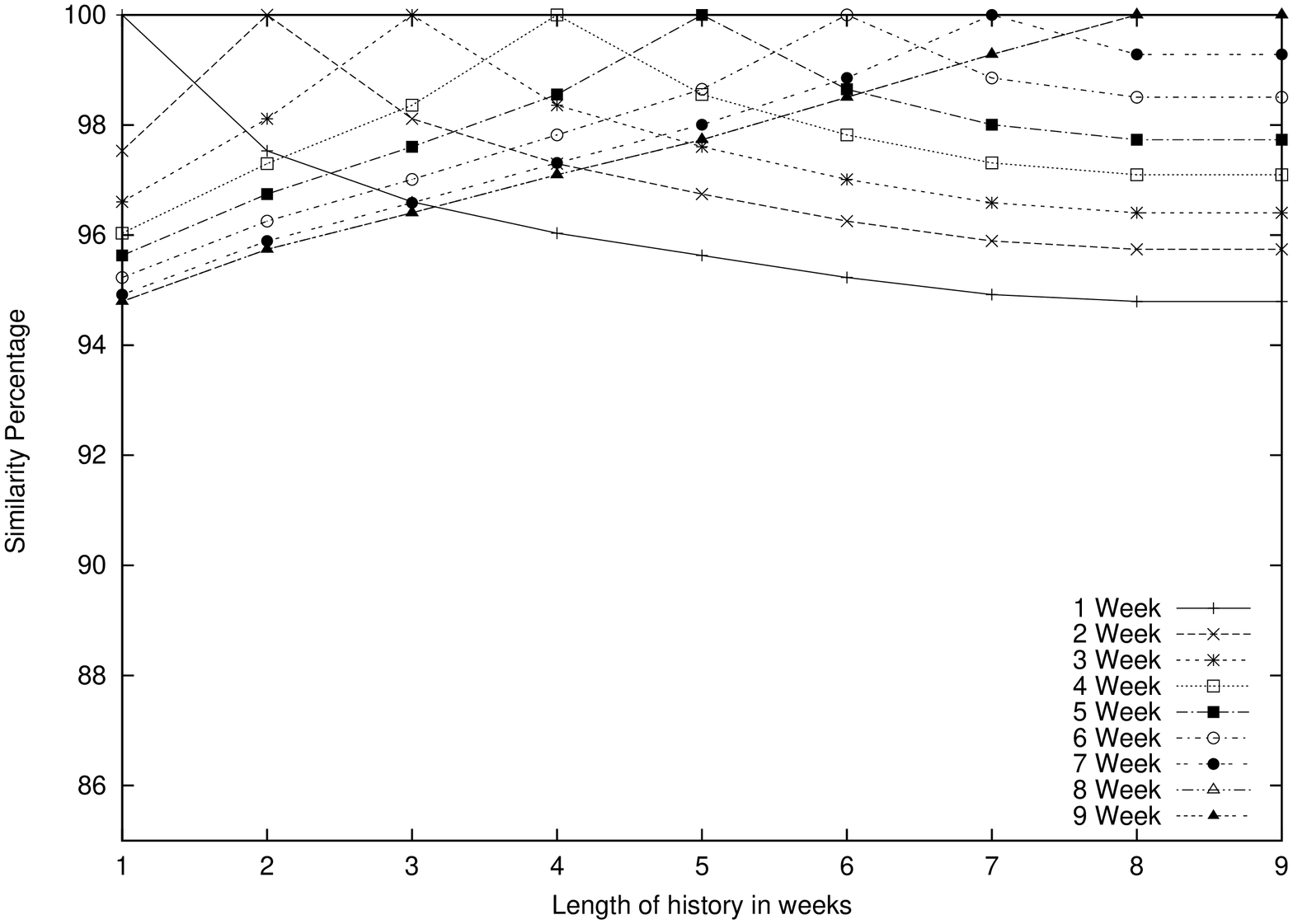, scale=0.22,angle=270}}
      \subfigure[Behavior Vector using Duration ($BV-D$)]{\epsfig{file=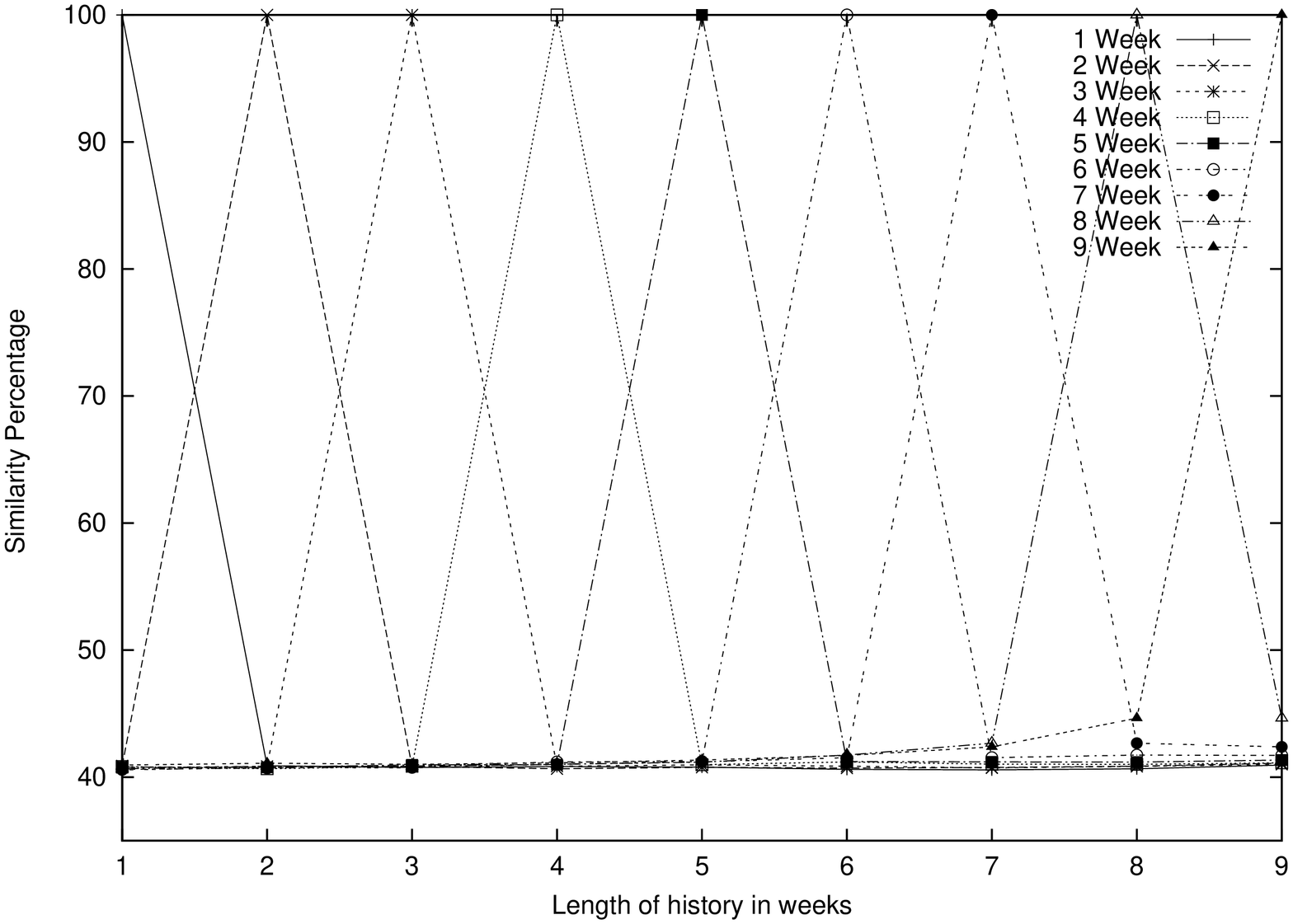, scale=0.22,angle=270}} 
      \subfigure[Behavior Matrix ($BM$)]{\epsfig{file=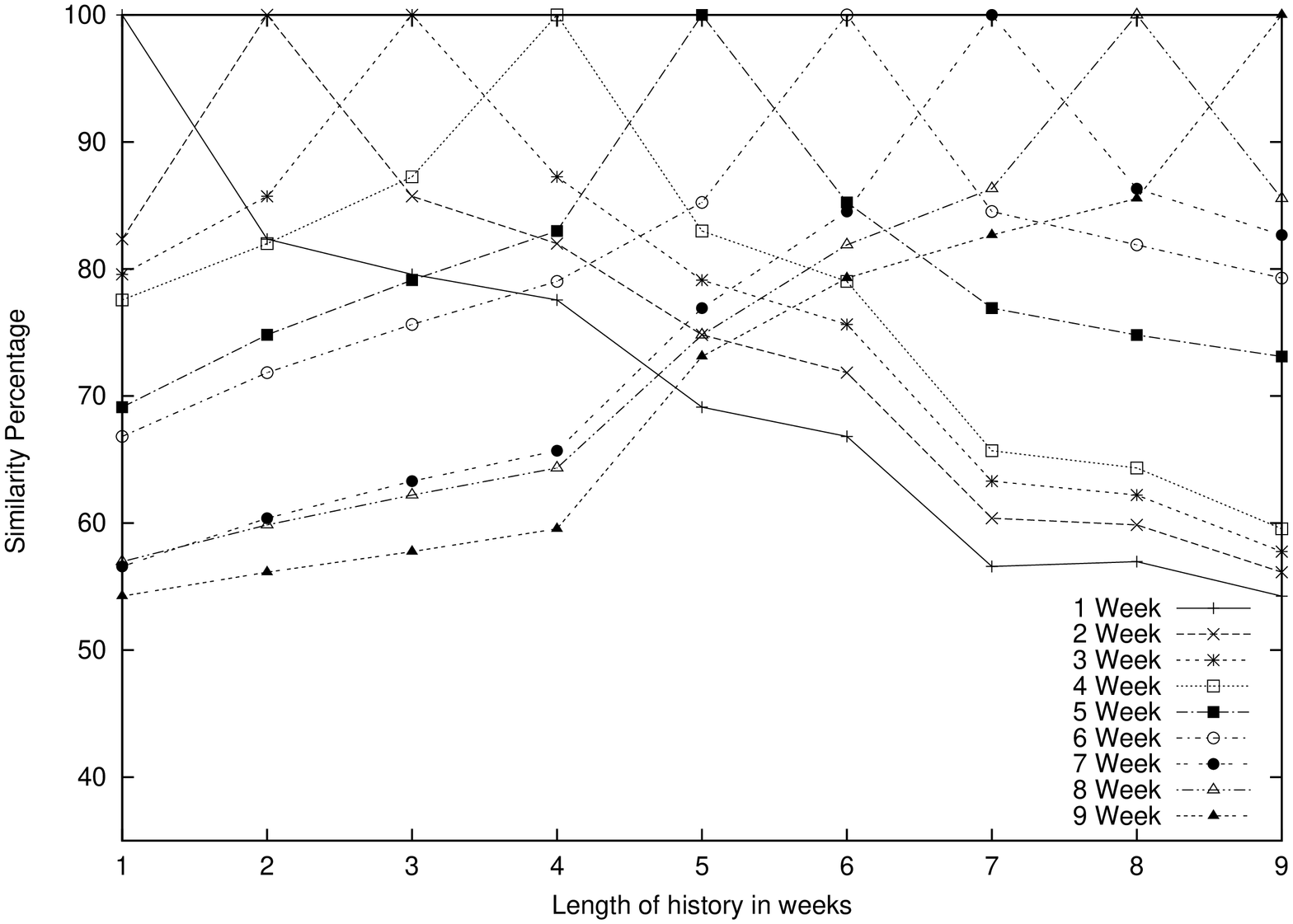, scale=0.22,angle=270}} 

    \caption{Comparison of trust list belonging to different history for various filters at T=40\% (note that the y-axis scale for $DE$, $FE$, and $BV-C$ starts at 85\% and for  $BV-D$ and $BM$ the scale starts at 35\%)}
    \label{fig:stability}
  \end{center}
\end{figure*}

How the trust list grows and evolves over time can provide insight into the stability of the trust list. One can also decide based on the stability how much history to have before one gets suitable recommendations. For this analysis, we compare the list of trusted nodes at various lengths of history with trust ($T$) value set at 40\%. Fig.~\ref{fig:stability} shows the stability of the various trust filters. We use 9 weeks long WLAN traces (Sep to Nov 2007) for the analysis. Except for the $BV-D$ metrics, all the other metrics show high similarity with trusted list of previous months. For $DE$, $FE$, and $BV-C$ the similarity between list for 1 week and 9 weeks is more than $89\%$. This implies that nodes trusted by $DE$, $FE$, and $BV-C$ using 1 week of history may still be trusted when considering 9 weeks of history with the probability greater than 89\%. This also implies that $DE$, $FE$, and $BV-C$ produce stable lists. This stability can be attributed to the fact that user encounters patterns captured by these three filters are inherently stable. 

$BM$ filter shows high stability in terms of trust list when the difference in history is less than 2 weeks (\textgreater 80\%) and the stability goes down to 55\% for the 1 week of history when compared to 9 week of history. 

For the $BV-D$ filter, the similarity between any list is around 40\%. This implies that every week the list changes by 60\%. This may point to a possibility that users go to same places regularly (by stability in  $BV-C$) but may spend different amounts of time at the same place over the weeks. This can cause $BV-D$ filter to pick different users in different length of behavior history.

Overall, we note that some filters ($DE$, $FE$, and $BV-C$) stabilize in just one week of history. These filters can be a good choice for recommendations when trust history is short and the time interval between trust list generation can also be long. For $BV-D$ filter we would have to redo the trust list every week as the stability is comparatively low.

\subsection{Correlation Between filters}

\begin{figure}
\centering

\includegraphics[width=2.5in,angle=270]{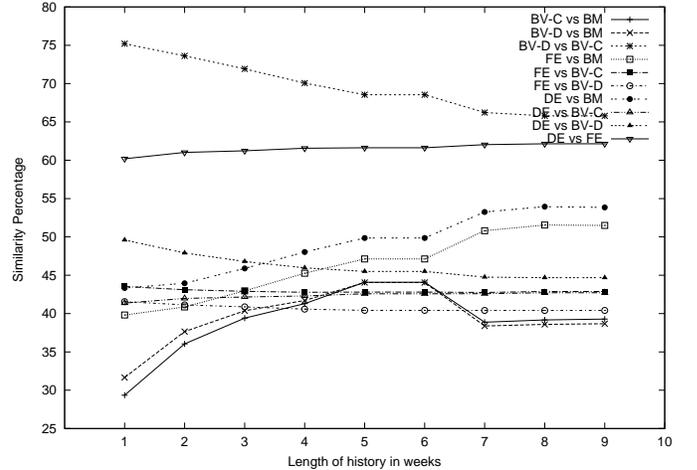} 
\caption{Comparison of similarity between trust lists generated by various filters at T=40\%}
\label{fig:correlate}
\end{figure}

It is important to study the differences between the trust list obtained from different filters over different length of traces (history). A smaller number of differences between a pair of filters would mean a stronger correlation between them and depending on the value it may make one filter redundant. On the other hand a large difference points to the orthogonality in the selection of users made by the filters. Fig.~\ref{fig:correlate} plots the results of comparing $DE$, $FE$, $BV-C$, $BV-D$,  and $BM$ filters. For these results, we have set trust at $T=40\%$. The results show comparison of top $T=40\%$ users given by the filters. 

Fig.~\ref{fig:correlate} also presents the effect of history of each filter on every other filter. For example, in $FE$ vs $DE$ curve, we compare lists produced by $FE$ filter using history of 1 to 9 weeks with the $DE$ filter.  We observe that $BV-C$ and $BV-D$ have the maximum similarity (\textgreater 65\%) and minimum similarity between the users is seen in the comparison of $BV-D$ (or $BV-C$) with $BM$.  We also notice that with time similarity between $FE$ (or $DE$) and $BM$ increases.

If two filters recommend different sets of trustworthy users, the user can be recommended with a richer set of users to trust. If in future, we use a self-learning recommendation system, based on previous selection of users, it can choose the filter that matches closely to the users choices. Similarly, if two filters present similar set of users, depending upon the space and memory requirements one of the filters may be removed.

\section{Effect of Trust and Selfishness} \label{epidemic}

In the previous sections we saw how trust can be developed in an infrastructure-less p2p network. In this section, we induce trust (using the filters) into a network with selfish nodes and analyze the effect on connectivity of the graph. The aim of this exercise is to study how trust can overcome disruption in connectivity of a network due to selfishness.

For this analysis we use Epidemic routing~\cite{epidemic}. Epidemic routing floods messages in time and space. The sender forwards a copy of the message to every encountered node that did not already receive the message. All these encountered nodes in turn forward a copy of the message to all the encountered nodes that did not already receive the message and so on.  Epidemic routing maximizes the message delivery rate and minimizes the message delay.  Epidemic routing in sense of message delivery success and message delay provides an upper bound on performance, i.e., no other routing protocol can provide better delivery rate or lesser delivery delay. This property allows us to use epidemic routing as a tool or benchmark to measure the dynamic connectivity of the encounter graphs with varying trust and selfishness.

Before coupling trust and selfishness, we look at the effects on the dynamic encounter graphs by the selfishness of the network nodes only. This analysis allows us to appreciate the effect of trust in the later part of this section. But before that we present the setup of our trace driven simulation.

\subsection{Trace-driven Simulation Setup}

For the simulation, we have used WLAN traces belonging to Sep, Oct, and Nov 2007. The epidemic routing is simulated only on the traces belonging to Nov 2007. Sep and Oct 2007 traces are used to develop the history database for various trust filters. The total number of users in Nov 2007 traces are 26,198. For epidemic routing, we have randomly selected 873 node as the sources (3.33\% of the total nodes) and all 26,198 nodes act as receivers. 

For analyzing each instance of the simulation, performance of epidemic routing is measured in terms of Unreachability, Delay and Overhead. We define \textit{Unreachability}  as the number of nodes out of all receivers (for Nov 2007 there are 26,198 receivers) that could not be reached  by a given source. \textit{Delay} is defined as the ratio of average time taken by a message to reach all the possible receivers over the max delay possible (for Nov 2007 it is 30 days). \textit{Overhead} is defined as the average number of hops a message took to reach all the possible receivers using the shortest path. In all the graphs, we have sorted the sources in the ascending order of the measured quantity. 

\subsection{Selfishness}
%need to introduce selfishness before this section. probably in introduction

In  infrastructure-less networks, without any incentive to route messages, nodes may become selfish and refuse to accept/route messages. Since an infrastructure-less network fundamentally depends on the participation of network nodes in routing and message transfers, selfishness can break down the connectivity in the network. 

To model the effect of selfishness on the connectivity of the network, we test the performance of epidemic routing when the network has selfish nodes. We define selfishness of a node as a probability ($S$) with which the node will not accept packet/message from another node for forwarding (assuming that if a node accepts packet for routing it would sincerely route it). For example if $S = 0.9$, then a node would not accept packets from 9 out of 10 nodes in the network. Using this probability ($S$) in the simulation of epidemic routing, with uniform random distribution, we decide the nodes to which a particular node is selfish to. (We have not considered misbehavior of the nodes but there are several techniques available to detect misbehaving nodes~\cite{Marti00mitigatingrouting,locationcentricisolation}. These techniques can be combined with our work.)

\begin{figure*}
  \begin{center}
\renewcommand{\thesubfigure}{\Alph{subfigure}.}

    \centering
    
      \subfigure[Unreachability]{\epsfig{file=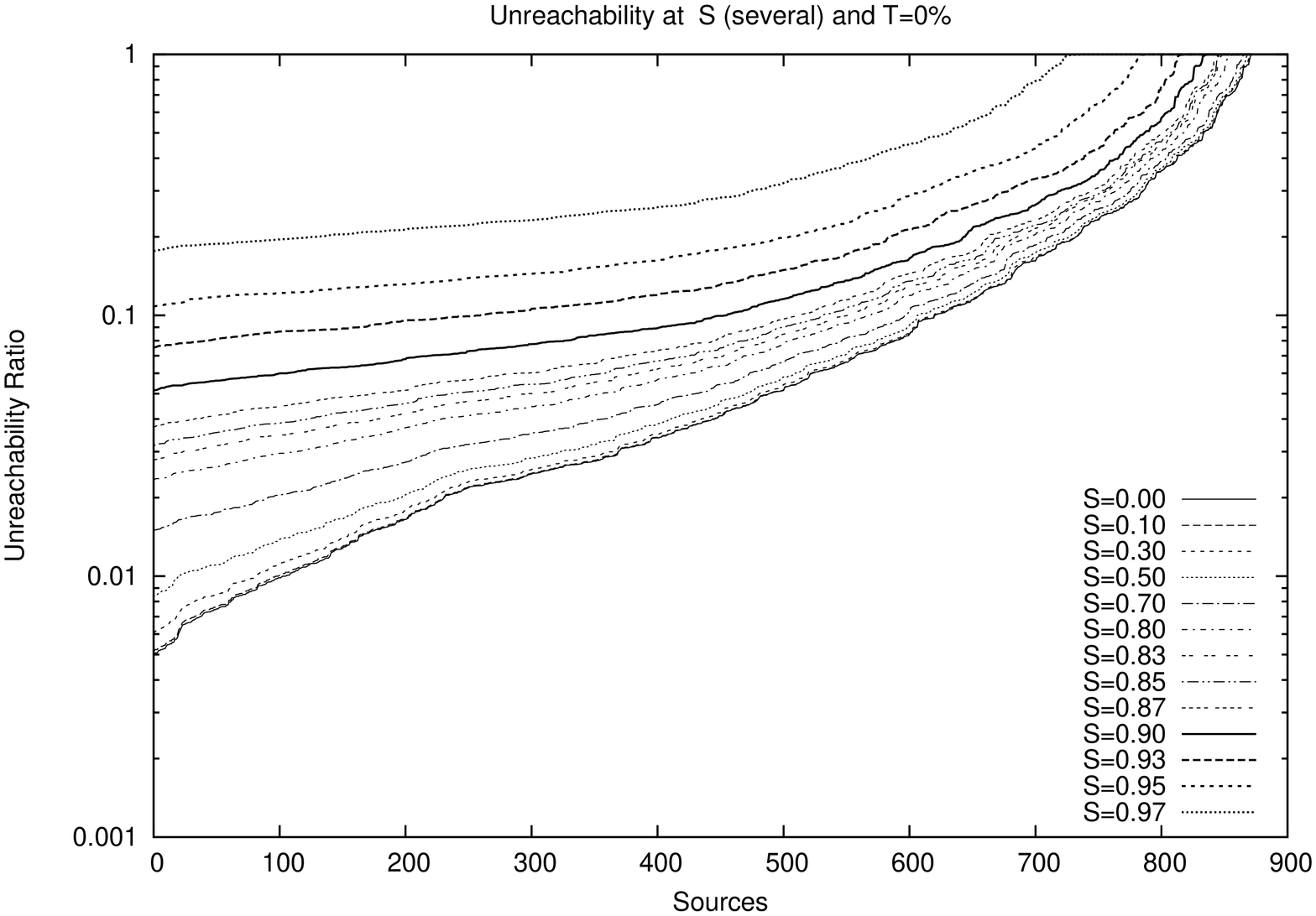, scale=0.23,angle=270}} 
      \subfigure[Delay]{\epsfig{file=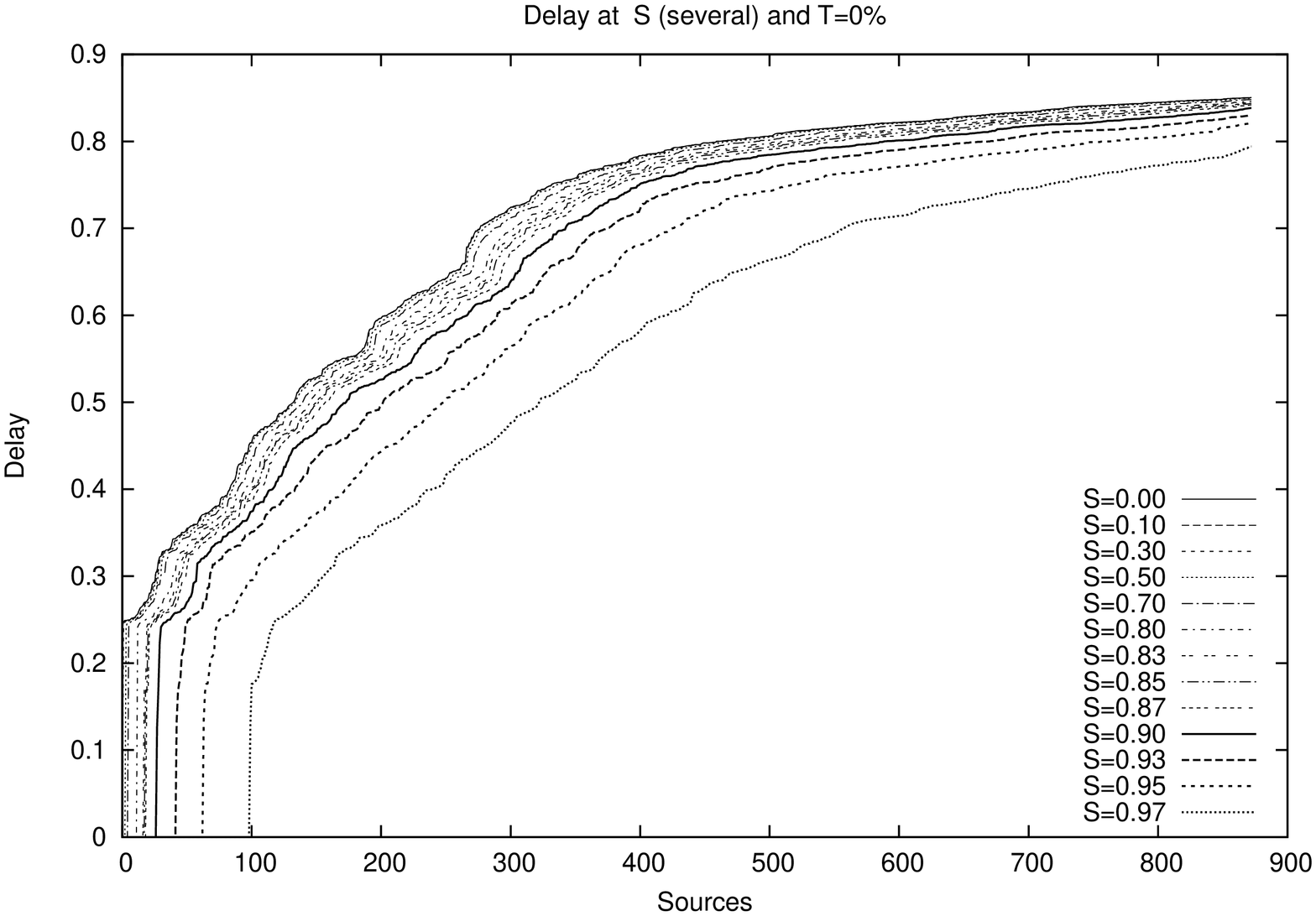, scale=0.23,angle=270}} 
      \subfigure[Overhead]{\epsfig{file=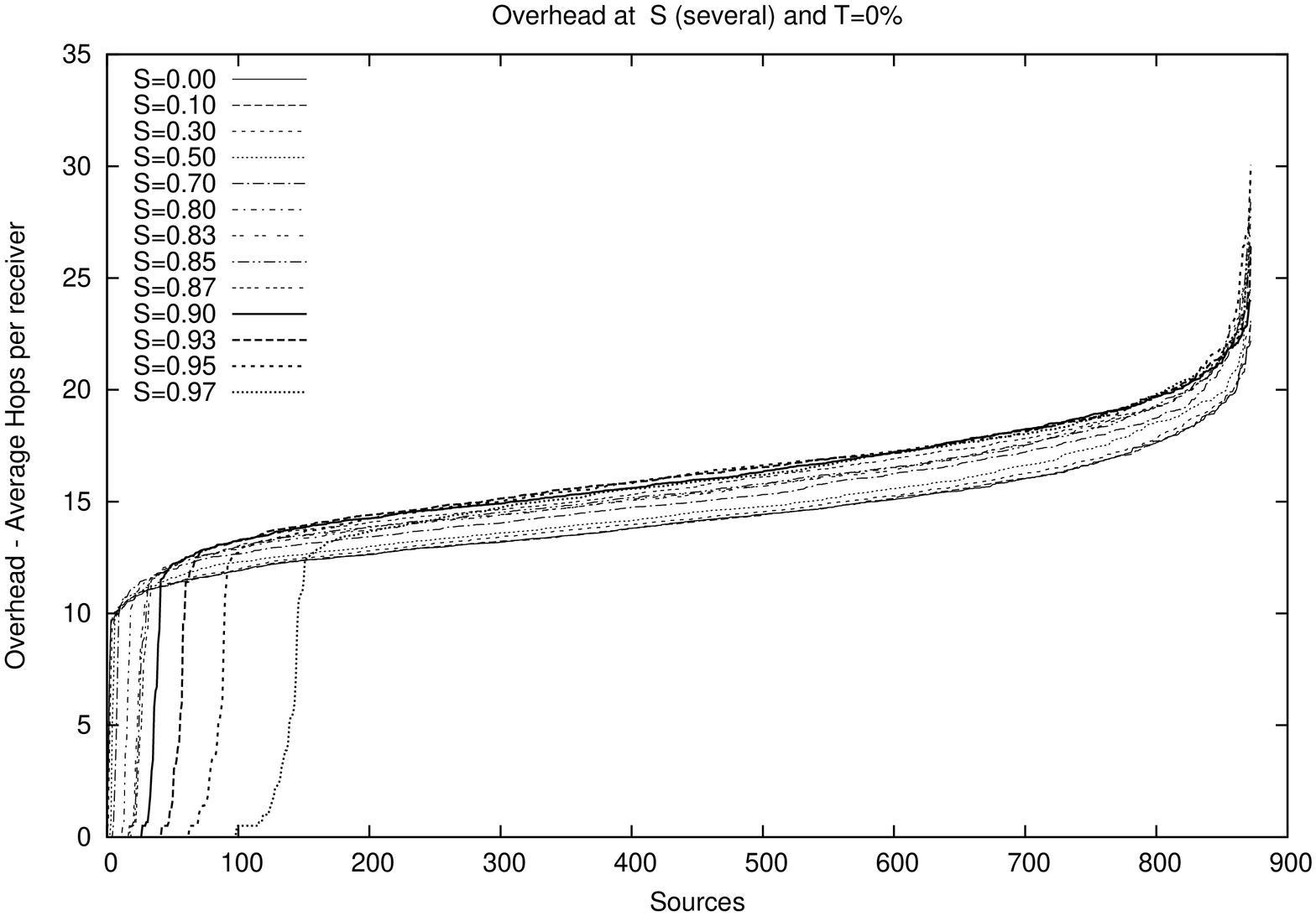, scale=0.23,angle=270}} 

    \caption{Unreachability, Delay  and Overhead at several S and T=0\%, during trace period of Nov 2007}
    \label{fig:varying_s}
  \end{center}
\end{figure*}

Fig.~\ref{fig:varying_s} shows the effect of selfishness with varying $S$ on the performance of epidemic routing in terms of unreachability, delay and overhead (hops per path). Looking at unreachability, we observe that it increases non-linearly  with increase in $S$ values. The average unreachability from $S=0$  to  $S=.97$ increases 4 times (1.6 times at $S=0.8$). Understanding delay requires us to understand how delay is calculated. Delay is only calculated for nodes that were reachable. So as the $S$ increases, less number of nodes are reachable and only these nodes are considered for delay calculations. Therefore we notice that delay decreases as $S$ increases because the reachability is decreasing. This is evident by the observation that as $S$ increases, increasing number of sources (12\% at $S=0.97$) report zero delay, indicating zero or very small reachability. Overhead measurement also reports overhead for the nodes that are reachable. We can see that with increase in $S$ the overhead increases as the messages have to take  more hops to reach a receiver. As the case with delay, in overhead too the increase in $S$ increases the number of sources which have very low hop count (mostly zero). This low count again points that several of sources (12\% at $S=0.97$) are unable to transfer messages to any receiver with increase in $S$.

The results show that selfishness in a network severely impacts the connectivity in the graphs. The increase in selfishness increases the network partitioning. More than 12\% of the users may not be able to transfer message to any user if $S>.97$ and unreachabilty increases 4 times.

\subsection{Effect of Trust}
%include comparison of various trust metrics.

Selfishness adversely impact connectivity in an encounter graph. To offset the impact, we propose the use of trust built upon interactions and similarity between the users. This section presents the impact of using trust on the connectivity in the encounter graph. We again use the epidemic routing as the tool for evaluation of network connectivity. 

Trust between a pair of user is generated using the trust filters developed in Sec.~\ref{trust_filter}. The filters select users based on homophily and interaction. To derive trusted users for the experiment we select top $T$ percent users based on the filters (For example, if $T=40\%$ and we are using $FE$ filter for a user A, then we would order all the users based on the number of encounters with A in decreasing order and select $40\%$ of those from the top as trusted users). In a practical implementation, users would have individual control on how many and which users to trust, but for the simulation, we assume that higher similarity (via trust filter) means greater homophily and thus greater trust between them.

\begin{figure}
\centering

\includegraphics[width=2in,angle=270]{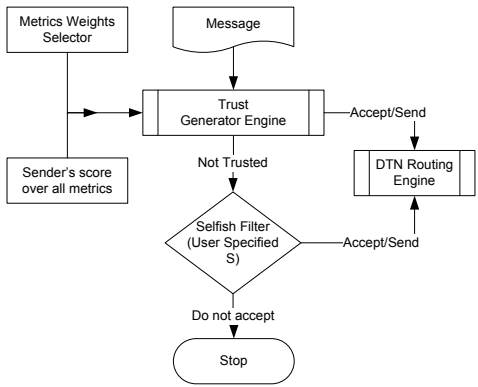} 

\caption{Flow chart for making routing decisions}

\label{fig:trust-gen}
\end{figure}

Once a user is considered trusted by a node, the node would accept all packets from this user and  attempt to route them, regardless of the selfishness ($S$). This condition is analogous to the fact that a user in real life may not be selfish to a person he/she knows as opposed to a stranger. Fig.~\ref{fig:trust-gen} shows the flow chart that a user's device may use before accepting and routing packets from another user. Here, an incoming message is processed by the Trust Generator Engine. This engine checks if the message is from a trusted node (information given by trust filters) and accepts the message if the node is trusted otherwise it is sent to the Selfish Filter. This selfishness filter stores user-specified selfishness probability and with this probability it decides whether to accept the message or reject it.

The trust lists are updated once a week, this can be justified (and is also realistic) as calculation of filter list after every encounter or even every day may occupy computational resources of the devices especially since many of the filters are stable. The comparison and stability with time is presented in previous section. For the simulation, filters bootstrap with two months of history and the trust lists are updated weekly.

\begin{figure*}
  \begin{center}
\renewcommand{\thesubfigure}{\Alph{subfigure}.}

    \centering
    
      \subfigure[Unreachability]{\label{fig:all_filters_unreachability}\epsfig{file=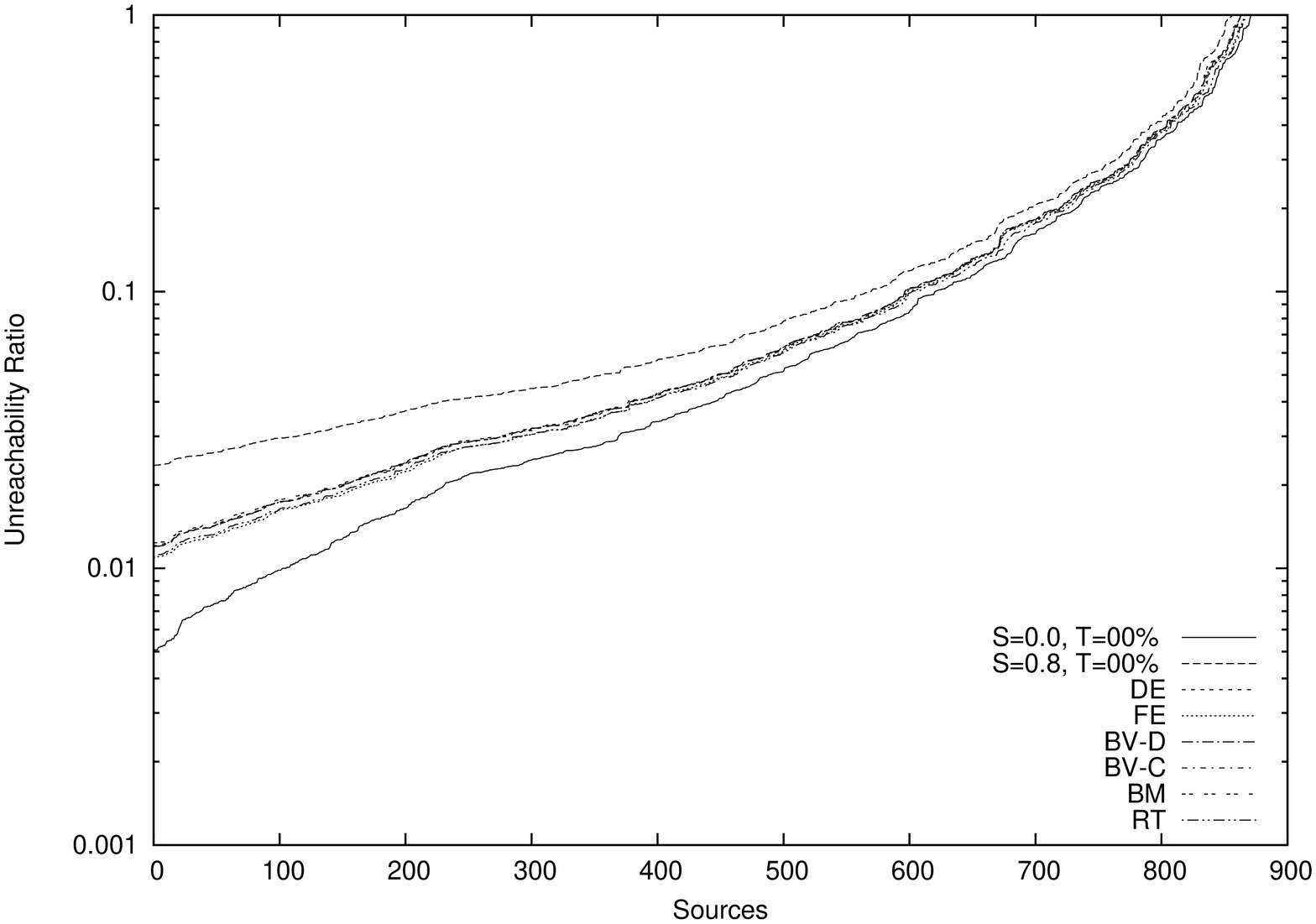, scale=0.23,angle=270}} 
      \subfigure[Delay]{\label{fig:all_filters_delay}\epsfig{file=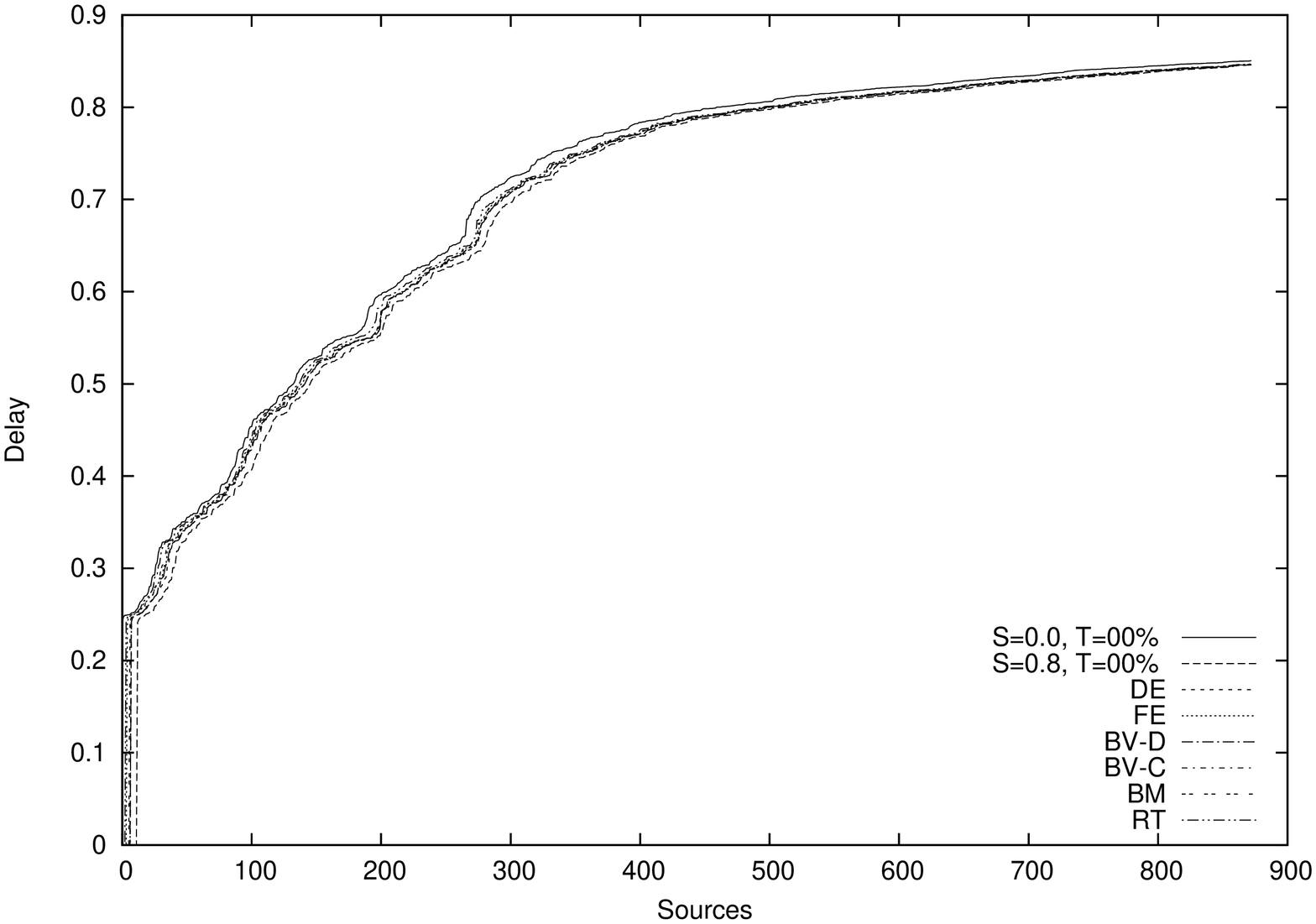, scale=0.23,angle=270}} 
      \subfigure[Overhead]{\label{fig:all_filters_overhead}\epsfig{file=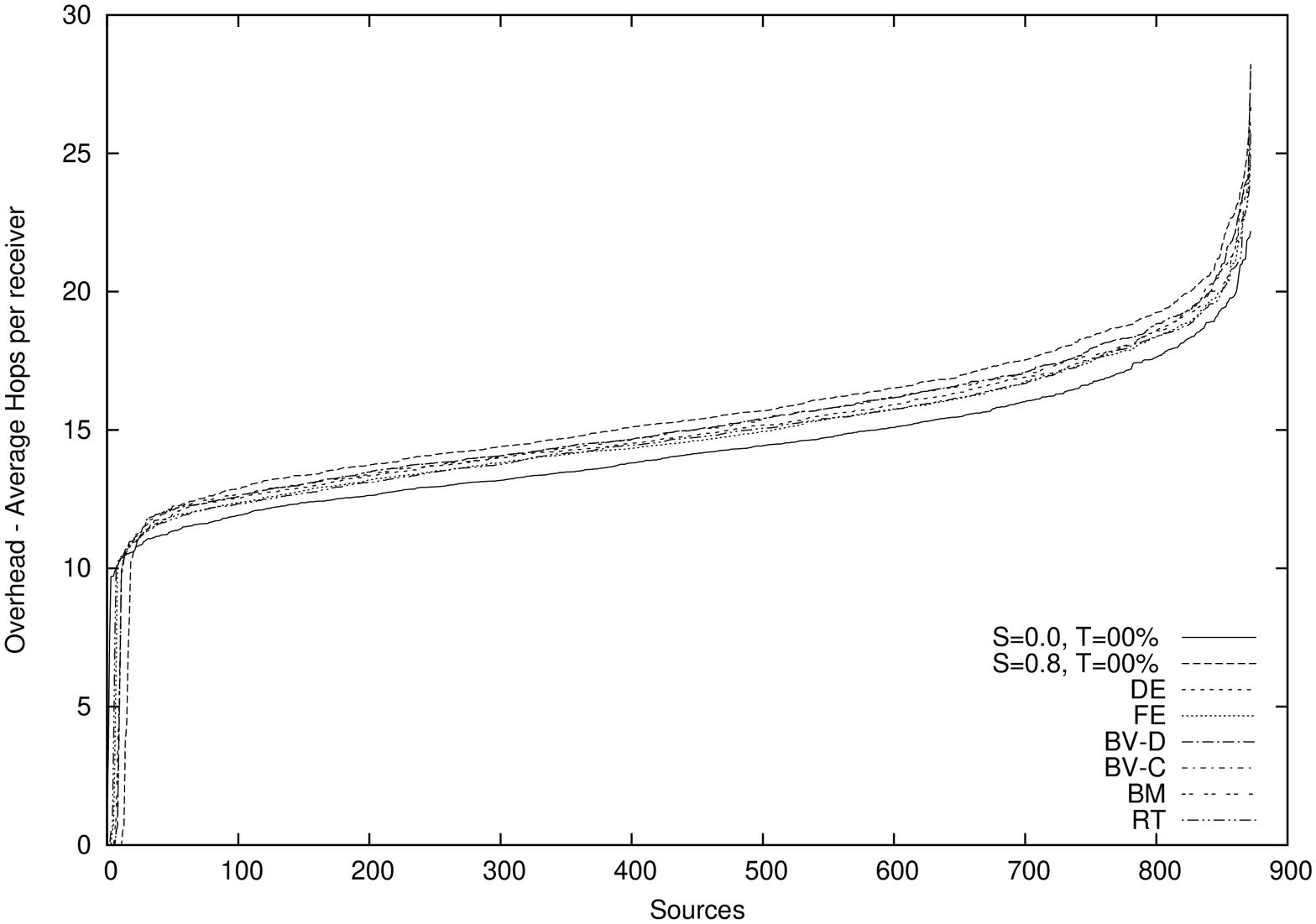, scale=0.23,angle=270}} 
     
    \caption{Unreachability, Delay and Overhead for all filters at S=0.8 and T=40\%}
    \label{fig:all_filters}
  \end{center}
\end{figure*}

 Fig.~\ref{fig:all_filters} shows the effect of trust on the performance (thus connectivity) when nodes are selfish. The trust value is fixed to $T=40\%$ and $S=0.8$. (We have performed several experiments like varying trust and history on all the filters, but due to lack of space we are presenting limited set of results.) We observe that the  presence of trust improves the performance of the network. The average unreachability goes down from $0.146$ to $0.122$ (from $S=0.8$ and $T=0\%$ to $S=0.8$ and $T=40\%$ ($FE$)), where as the average unreachability at $S=0.0$ and $T=0\%$ (no selfishness)  is $0.109$. The unreachability at $S=0.8$ and $T=0\%$ is $33\%$ more than the network with no selfishness, the trust reduces this unreachability to $11\%$ more than the network with no selfishness. We observe that RT filter performs marginally better in terms of improvements in the reachability. This may be because RT selects users randomly which may be providing shortcuts in the encounter graph. All the filters show similar performance for unreachability.

The Delay graph (Fig.~\ref{fig:all_filters_delay}) shows that the delay decreases with selfishness, this is because the connectivity goes down with selfishness and thus the time taken to reach the reachable nodes also decreases. So delay may not give correct information, but we can argue that adding trust is moving the delay towards the scenario where the network has no selfishness. The Overhead graph (Fig.~\ref{fig:all_filters_overhead}) shows trends similar to unreachability.  

All the filters provide similar performance. Tab.~\ref{tab:all_filter} provides average of unreachability, delay and overhead for all the filters. $FE$ filter performs the best (apart from RT filter). However, both $BV-C$ and $BV-D$ perform slightly inferior to $DE$ and $FE$. 

\begin{table}[htp]
\begin{center}
\begin{tabular}{c c c c}
\hline Filter & Unreachability & Delay & Overhead\\
\hline S=0.0 \& T=0\% &0.108998 &0.707754 &14.2889	\\
 S=0.8 \& T=0\% &0.145975&0.688934&15.4238\\
DE &0.125623&0.695938&14.9732\\
FE &0.121892&0.697582&14.8221\\
BV-D &0.125686&0.695215&15.1567\\
BV-C &0.125686&0.695215&15.1567\\
BM   & 0.12458 & 0.696189 & 15.0937\\
RT &0.119043&0.699930&14.8389\\

\hline \\
\end{tabular}\end{center} 
\caption{Average Unreachability, Delay and Overhead using various filters (S=0.8 \& T=40\%)}
\label{tab:all_filter}
\end{table}

\subsection{Trust and Selfishness}

\begin{figure*}[htp]
  \begin{center}
\renewcommand{\thesubfigure}{\Alph{subfigure}.}

    \centering
    
      \subfigure[Unreachability]{\epsfig{file=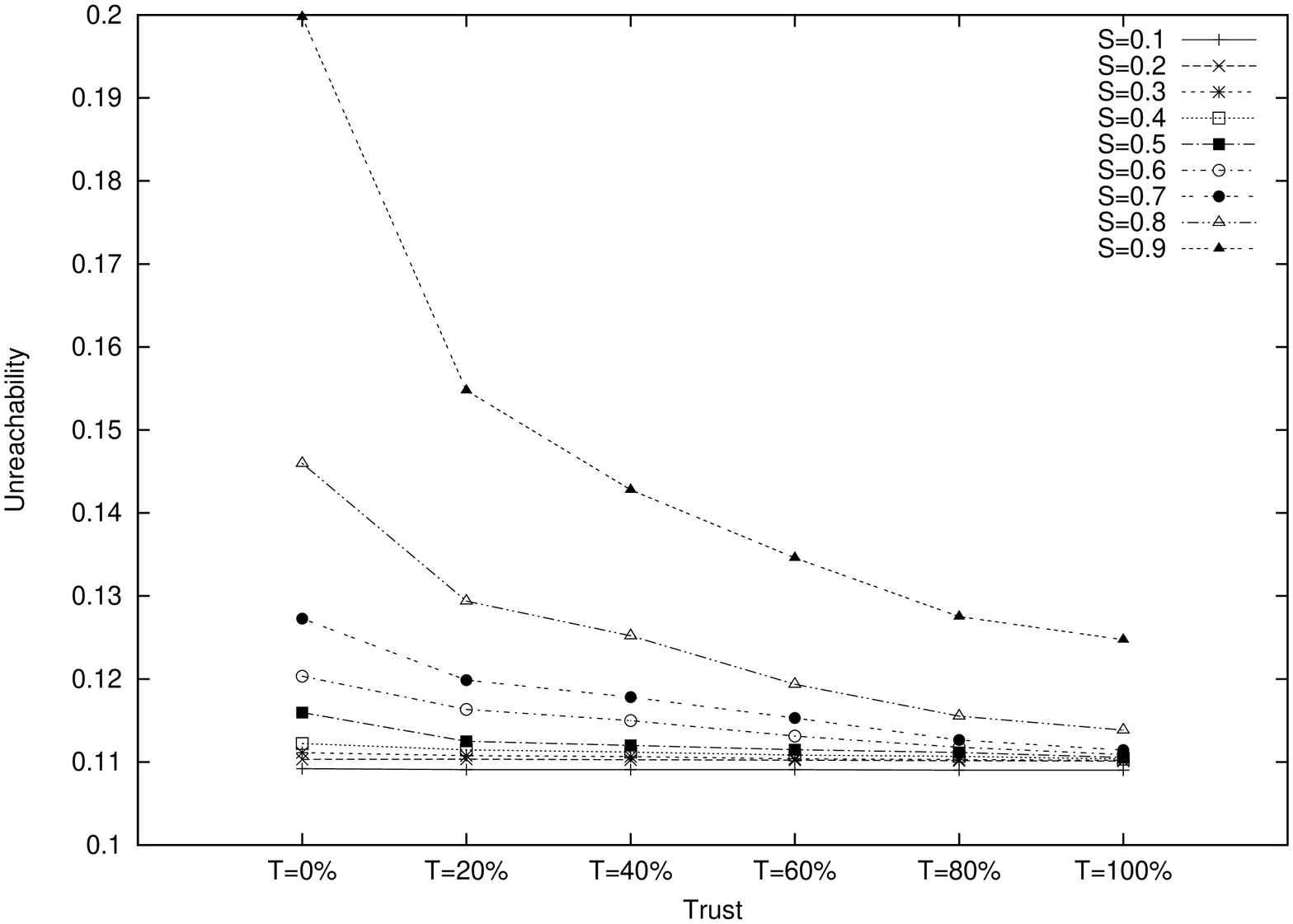, scale=0.23,angle=270}} 
      \subfigure[Delay]{\epsfig{file=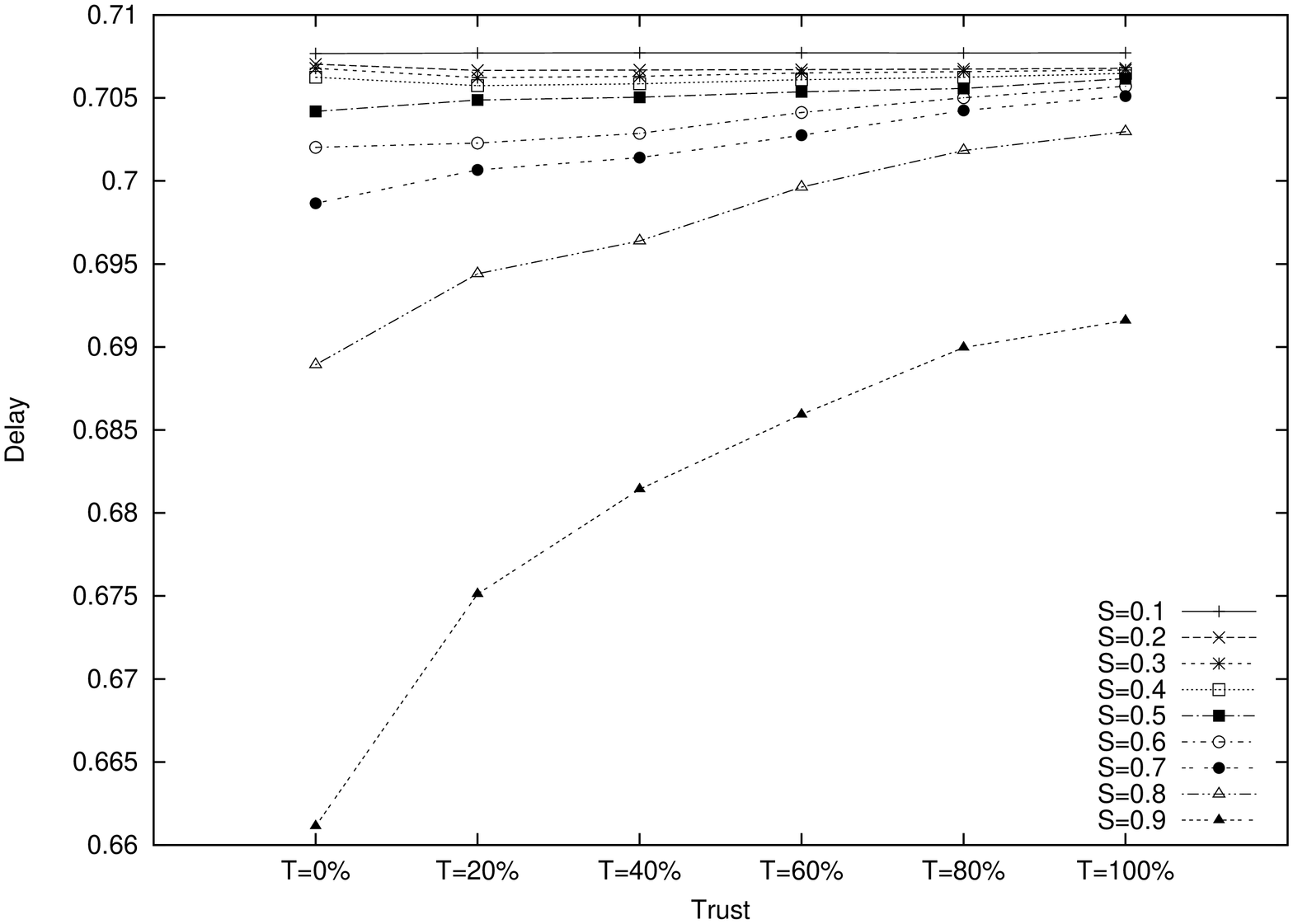, scale=0.23,angle=270}} 
      \subfigure[Overhead]{\epsfig{file=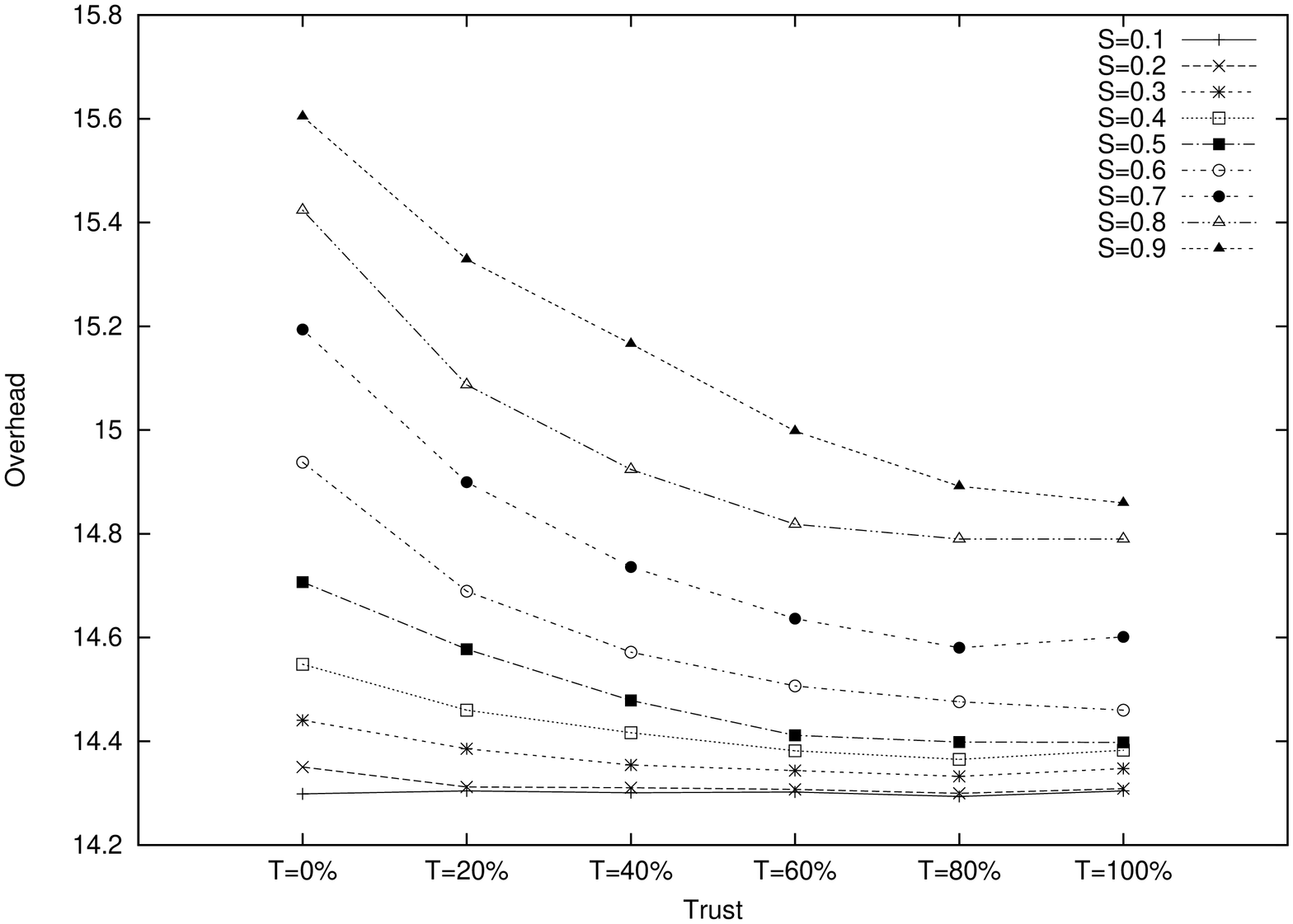, scale=0.23,angle=270}} 
     
    \caption{Average Unreachability, Delay and Overhead at various values of T and S for $DE$ filter}
    \label{fig:T_vs_S}
  \end{center}	
\end{figure*}

To understand the effect of selfishness and trust on each other, Fig.~\ref{fig:T_vs_S} shows average unreachability, delay and overhead at multiple $S$ and $T$ values using $DE$ filter (other filters give similar results). We observe that increase in trust ($T$) reduces unreachability, delay and overhead regardless of selfishness ($S$). The effect of trust is higher when the selfishness in the network is high (the gradient of $S=0.9$ is maximum between any two values of trust). This is a critical result since the utility of trust increases with the increase in selfishness of the network. \textit{The more is the selfishness, the more is the value derived from trust.} Similarly, we notice that for  $S=0.1$  to $0.4$ (a low value of selfishness) where increase in  trust does not affect the connectivity in the network in a major way. So, trust is useful mainly in the networks with high selfishness ($S > 0.4$). We also notice that maximum change happens in unreachability, delay and overhead at $T=20\%$ from $T=0\%$. 
For this reason we have selected $S=0.8$ in the previous experiment as it serves as  a middle ground.

%\begin{figure*}[htp]
%  \begin{center}
%\renewcommand{\thesubfigure}{\Alph{subfigure}.}
%
%    \centering
%    
%      \subfigure[Unreachability]{\epsfig{file=List-dura-Unreachability.ps, scale=0.23,angle=270}} 
%      \subfigure[Delay]{\epsfig{file=List-dura-Delay.ps, scale=0.23,angle=270}} 
%      \subfigure[Overhead]{\epsfig{file=List-dura-Overhead.ps, scale=0.23,angle=270}} 
%     
%    \caption{Unreachability, Delay and Overhead at various values of T and S=0.8 for $DE$ filter}
%    \label{fig:DE_t_s}
%  \end{center}
%\end{figure*}

In this section we observed how selfishness can affect the connectivity in the encounter graphs. We also observed that using the trust list created by the trust advisor filters the effects of selfishness can be reduced. 
%\section{Applications}

%\section{Implementation}

\section{Conclusions and future Work} \label{future}

In this work we discuss novel methods that can convert social interactions and similarities into trust for cooperation in mobile adhoc and DTN. We propose several trust advisor filters that can measure similarity between two users. We perform statistical studies to capture the properties of the filters. We find that several filters produce trust list that are stable for more than 2 months. We note that results of different filters are noticeably different  and  thus are capable of providing a rich selection to the user.  Our paper  also examines effects on network connectivity due to selfishness in the network using real WLAN traces. Selfishness adversely affects the reachability and partitions the connectivity graph. Unreachability increases by 150\% as the selfishness reaches $S=0.8$. We also analyze the effects of having trusted nodes in a selfish network. We note that having trusted nodes increases the connectivity and thus the reachability in the network with selfish nodes. At trust of $T=40\%$ in a selfish network ($S=0.8$) the unreachability remains at 120\% of the levels of a network with no selfishness. We note that in a selfish network if the nodes trust the nodes that are similar to themselves, the effect of selfishness in the network can be drastically reduced. 

A major application of this work is to serve as a guide to applications running on  a mobile devices in making routing and forwarding decisions, where  presence of a node in the trusted list can increases the chances of message delivery.

In the future, we plan to work on analyzing other trust filters. We would also like to implement a scoring/weight system that can combine together the scores of all the filter for a encountered user. This score system would allow a user to consider single score before adding another user to the trust list. We would also like to design protocols that utilize the filters and implement these filters on a mobile device.

% For other papers future work
%\subsection{Other filters}
%The set of filters we have considered above can handle several primary scenarios for capturing similarity. These filters are not the only filters that can be used, we list several more but due to limitation of space and time we have not explored them in detail. 
%
%\begin{enumerate}
%\item{Inter-contact Time b/w Encounters:}  Time between two encounters of a user with a particular user. 
%\item{Encounters when other friends are present:} Number of friends encountered at the time of encountering this particular user.
%\item{Location of Encounters relative to user's top location:} Yet another way to measure similarity if encounters are happening in the location preferred by the user.
%\end{enumerate}

\bibliographystyle{IEEEtran}	% (uses file "plain.bst")
\bibliography{../../mybib}
\end{document}